\documentclass[sigconf, nonacm, screen]{acmart}
\AtBeginDocument{%
  }

\makeatother

\renewcommand\footnotetextcopyrightpermission[1]{} 
\setcopyright{none} 

\usepackage{enumitem}
\usepackage{color,soul}
\usepackage{todonotes}
\usepackage{booktabs}    
\usepackage{tabularx}    
\usepackage{array}       
\begin{document}

\title{LLM Use for Mental Health: Crowdsourcing Users' Sentiment-based Perspectives and Values from Social Discussions}

\settopmatter{authorsperrow=4}
\author{Lingyao Li}
\authornote{These authors contribute equally to this research and are corresponding authors.}
\email{lingyaol@usf.edu}
\affiliation{%
  \institution{University of South Florida}
  \city{Tampa}
  \country{USA}
}

\author{Xiaoshan Huang}
\authornotemark[1]
\email{xiaoshan.huang@mail.mcgill.ca}
\affiliation{%
  \institution{McGill University}
  \city{Montreal}
  \country{Canada}
}

\author{Renkai Ma}
\authornotemark[1]
\email{renkai.ma@uc.edu}
\affiliation{%
  \institution{University of Cincinnati}
  \city{Cincinnati}
  \country{USA}}

\author{Ben Zefeng Zhang}
\email{ben.z.zhang@stonybrook.edu}
\affiliation{%
  \institution{Stony Brook University}
  \city{Stony Brook}
  \country{USA}
}

\author{Haolun Wu}
\email{haolun.wu@mail.mcgill.ca}
\affiliation{%
 \institution{McGill University}
 \city{Montreal}
 \country{Canada}}

\author{Fan Yang}
\email{YANG259@mailbox.sc.edu}
\affiliation{%
  \institution{University of South Carolina}
  \city{Columbia}
  \country{USA}}

\author{Chen Chen}
\email{chechen@fiu.edu}
\affiliation{%
  \institution{Florida International University}
  \city{Miami}
  \country{USA}}


\newcommand{\red}[1]{\textcolor{red}{#1}}



\begin{abstract}
Large language models (LLMs) chatbots like ChatGPT are increasingly used for mental health support. They offer accessible, therapeutic support but also raise concerns about misinformation, over-reliance, and risks in high-stakes contexts of mental health. We crowdsource large-scale users' posts from six major social media platforms to examine how people discuss their interactions with LLM chatbots across different mental health conditions. Through an LLM-assisted pipeline grounded in Value-Sensitive Design (VSD), we mapped the relationships across user-reported sentiments, mental health conditions, perspectives, and values. Our results reveal that the use of LLM chatbots is condition-specific. Users with neurodivergent conditions (e.g., ADHD, ASD) report strong positive sentiments and instrumental or appraisal support, whereas higher-risk disorders (e.g., schizophrenia, bipolar disorder) show more negative sentiments. We further uncover how user perspectives co-occur with underlying values, such as identity, autonomy, and privacy. Finally, we discuss shifting from ``one-size-fits-all'' chatbot design toward condition-specific, value-sensitive LLM design.
\end{abstract}



\maketitle

\section{Introduction}


Large Language Model (LLM) chatbots like ChatGPT~\cite{ChatGPT} have quickly become part of our everyday life, with millions turning to them for information, conversation, and, increasingly, for mental health support~\cite{xu2024mental,badawi2025can}. Both service providers and users often describe these chatbots as accessible, always-available, and non-judgmental listeners that facilitate their therapeutic needs or personal reflection, particularly when professional mental health care is too difficult to access or afford~\cite{Lawrence2024, Hua2025}. These advantages highlight the potential of LLMs to supplement mental health support.

However, pressing concerns have emerged. LLMs often lack contextual understanding, leading to inappropriate or misleading advice, masked by their  persuasive tone~\cite{Guo2024}. These risks are especially salient in recent incidents, where users experienced emotional distress, self-harm, and even suicide, after interacting with chatbots that reinforced harmful thoughts~\cite{Dupre2025, Gold2025, Klee2025}. These underscore the dangers of over-reliance and highlight the ethical need for human values \cite{friedman2013value} to be incorporated into LLM chatbots, such as privacy and accountability, particularly in sensitive mental health contexts.

A growing body of work has probed these opportunities and risks and has shown that LLM chatbots can deliver mental health education, assessment, intervention, and empathetic reflections that users perceive as calming or supportive~\cite{Schmidmaier2025a, Schmidmaier2025b, Lai2023}. Benchmarking and audit studies~\cite{xu2024mental, Patil2025} further demonstrate that alignment-tuned models produce safer, more compassionate responses than earlier generative systems. Meanwhile, interview- and survey-based research has explored self-disclosure, help-seeking behavior, and emotional regulation in the context of LLM use~\cite{Muller2024, kwesi2025exploring}. 

While existing evidence collectively shows that LLMs hold promise for mental health support, it remains unclear how users’ sentiments (\emph{emotional attitudes}) and perspectives (\emph{rational opinions}) shape and are shaped by their interactions with chatbots and the values involved. For example, how are LLMs used for mental health management? Under what mental health conditions do their perspectives emerge? And how are the values in LLM design associated with users' LLM chatbot use across mental health conditions? Understanding these requires analyzing large-scale data from popular approaches such as crowdsourcing through social media or surveys to capture users' self-report experiences with LLM use~\cite{li2024chatgpt, li2025towards, wise2025crowdsourced}.

To address these challenges, we leverage crowdsourced data from multiple social media platforms to examine how users discuss and experience the intersection with LLM chatbots regarding mental health. 
Using an LLM-assisted analytical pipeline, we extract and categorize social discussion posts that mention major LLM chatbots and mental health conditions such as depression, anxiety, and ADHD. Grounded in the principle of Value-Sensitive Design (VSD)~\cite{friedman2013value}, our study is guided by two research questions (RQs):

\begin{itemize}[leftmargin=*]
\item \textbf{RQ1:} How do users' sentiments toward LLM chatbots use vary across different mental health conditions?
\item \textbf{RQ2:} How are mental health conditions associated with users’ perspectives and values in their interactions with LLM chatbots?
\end{itemize}
\vspace{-0.3mm}

Our contributions are threefold. First, we provide a large-scale, cross-platform empirical mapping of real-world discussions linking LLMs and mental health. Second, we construct an impact typology that delineates the conditions under which LLMs appear to benefit versus harm users’ emotional well-being. Third, we offer design implications for developing safer, context-sensitive AI systems and guiding stakeholders in responsible LLM use for mental health. Overall, our analysis offers an evidence-based foundation for understanding LLMs’ bidirectional impacts on mental health in the real-world contexts.

\section{Related Work}

\subsection{LLMs and Mental Health}%
%

\noindent{\bf Supporting mental health through LLM chatbots.}
Prior research has explored how AI-driven chatbots and LLM can be used to support mental health education, assessment, and invention~\cite{Lawrence2024, wang2025application, Hua2025, vaidyam2019chatbots,feng2025effectiveness, li2023systematic}.
For example, PSY-LLM~\cite{Lai2023} and PsycoLLM~\cite{Hu2025PsychoLLM} showed how LLM-based chatbots can help alleviate demand on psychological counseling services.
Song et al.~\cite{song2025typing} pointed out that users often value these chatbots for their perceived non-judgmental nature.
Sharma et al.~\cite{sharma2024facilitating} demonstrated how LLM can support self-guided mental health interventions through cognitive restructuring, an evidence-based therapeutic technique to overcome negative thinking.
Using mobile sensor data and an edge LLM, MindGuard~\cite{Ji2024MindGuard} demonstrated how ecological momentary assessment can be designed for mental health screening and intervention conversations.
SouLLMate~\cite{Guo2025SouLLMate} demonstrated an adaptive LLM-driven system for suicide risk detection and proactive guidance dialog.
Conversely, researchers have noted some concerns.
For example, recent studies highlighted the negative impacts of hallucinations, misleading advice, and clinically ungrounded albeit persuasive responses~\cite{jin2025applications,hipgrave2025balancing}. 
Moreover, anthropomorphism and relational bonding may intensify over-reliance, shaping users’ emotional interpretations, social judgments, and decision-making in ways that may not always align with users' well-being~\cite{glickman2025human,jairoun2024benefit}.

\vspace{4px}\noindent{\textbf{Impact of LLM use on mental health.}}
A second body of work has explored the impact of LLM interactions on mental health.
Fang et al.~\cite{fang2025ai} found that higher daily interaction with chatbots correlates with users' increased emotional dependence and reduced real-world socialization \cite{fang2025ai}. 
While existing research such as \cite{song2025typing, li2025design, sharma2024facilitating} has focused on individual mental health conditions related to social isolation, there is a lack of comprehensive, cross-condition understanding of how user sentiments, perceived benefits, risks, and value-related considerations vary across the spectrum of mental health conditions. Our study directly addresses this gap by offering a multi-condition, multi-LLM perspective on users discussing and experiencing LLM chatbots for mental health support.

\vspace{4px}\noindent{\textbf{Use LLMs to understand mental health through online texts.}}
Closely related to our approach, another line of work has explored insights into mental health through social media and online texts~\cite{Choudhury2014, Guntuku2017, Choudhury2013, Choudhury2013b}. Later, Mental-LLM~\cite{xu2024mental} examined the feasibility of using LLMs to perform various mental health prediction tasks based on online text. MentaLLaMA~\cite{Kailai2024MentaLLaMA} and Cognitive-Mental-LLM~\cite{Patil2025} further explored using LLMs for interpretable mental health analysis on social media. Zhao et al.~\cite{Zhao2025} designed an LLM-based topic modeling framework to analyze public discussions about mental health on social media. Unlike prior research, our goal is to understand users’ sentiments, perspectives, and values associated with their interactions with LLMs in the context of mental health, grounded in the VSD framework and informed by both LLM-assisted information extraction and qualitative content analysis.

\vspace{-3mm}
\subsection{Achieving Value-Sensitivity}

Designing interactions with LLM chatbots requires a framework that incorporates \emph{human values}, extending beyond traditional usability metrics \cite{Nielsen1993UsabilityEngineering}.
One relevant and widely adopted framework is VSD, in which \emph{human values} are understood as ethically important principles that promote wellbeing, empathy, dignity, and experience~\cite{friedman2013value}.
In a nutshell, VSD emphasizes that technology is not value-neutral but both \emph{shapes} and \emph{is shaped} by the value of its stakeholders~\cite{friedman2013value}. 
Translating VSD principles into LLM-powered applications presents significant challenges; unlike traditional software, where values can be deterministically encoded, LLMs represent human values probabilistically, learned from large datasets, which may not always align with user needs \cite{liscio2025value, sadek2024guidelines}.

Applying VSD to LLM-powered applications in the context of mental health allows researchers to examine how human values manifest in practice and influence on user sentiments and perspectives.
This empirical grounding can, for example, help preserve the ``contextual integrity'' of user interactions \cite{nissenbaum2011contextual}, as current ``one-size-fits-all'' safety protocols of LLM chatbots often fail to account for the privacy and information disclosure norms \cite{ali2025understanding, li2025towards}. 
By synthesizing these user-centered insights with technical design processes, a VSD-informed analysis can guide LLM chatbot development toward participatory value alignment, ensuring that future chatbots are responsive to the context-sensitive values of users \cite{guan2025lived} 
%
also clinically mindful.

\vspace{-1mm}
\section{Methods}

\begin{figure*}[t]
\centering
\includegraphics[width=0.92\textwidth]{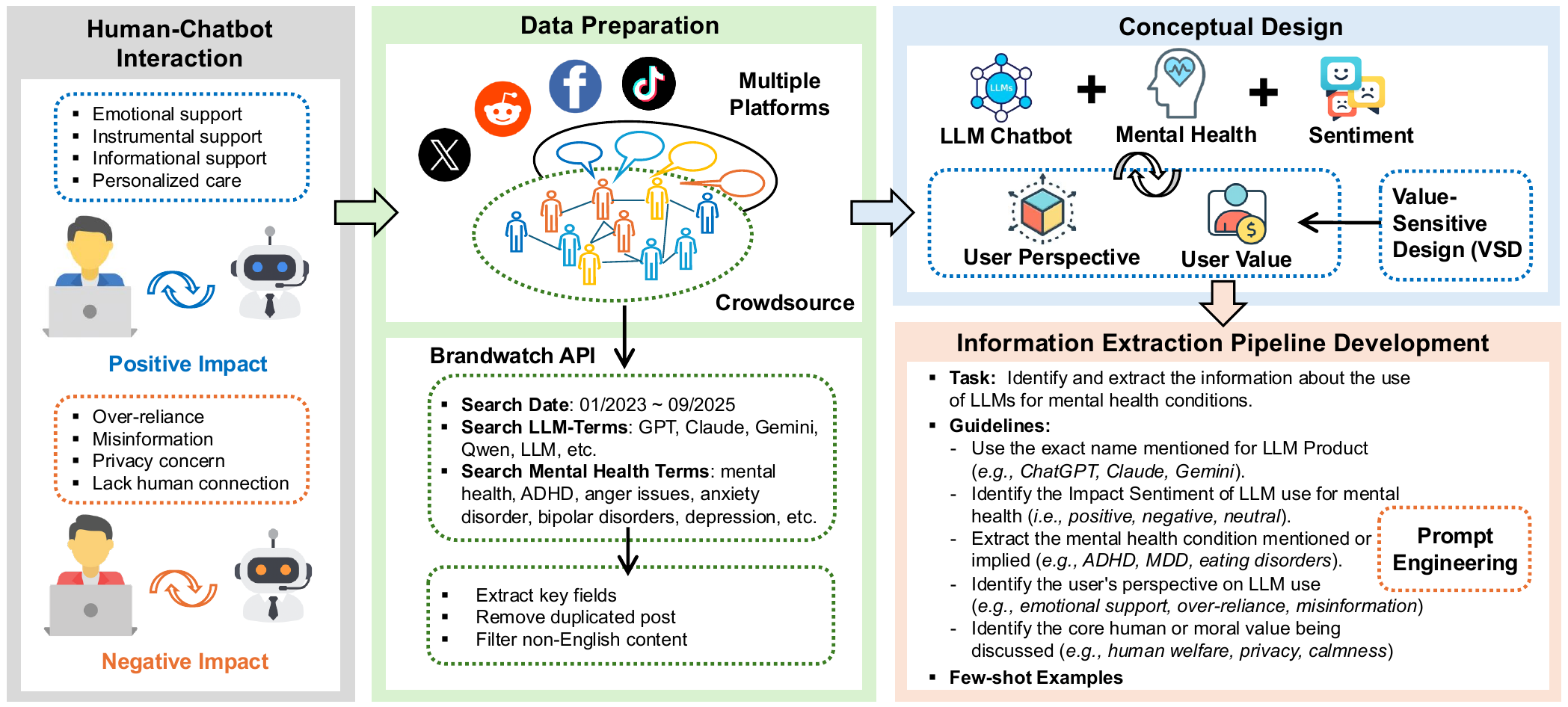}
\vspace{-0.1in}
\caption{Research framework to implement data preparation and methods.}
\vspace{-0.1in}
\label{fig:framework}
\end{figure*}

Figure~\ref{fig:framework} provides an overview of our research framework, which integrates data collection, conceptual design, and an LLM-assisted extraction pipeline. We first crowdsource posts referencing both LLMs and mental health conditions across major social media platforms. Our conceptual design draws on human–chatbot interaction research and VSD~\cite{friedman2013value,friedman2019value} to define the key constructs extracted from each post, including LLM chatbot, impact sentiment, mental health condition, user perspective, and user value. Guided by this schema, we develop a structured extraction pipeline that combines prompt engineering, few-shot examples, and chain-of-thought reasoning to identify multi-dimensional information from each post.

\vspace{-1mm}
\subsection{Data Preparation}

We use Brandwatch\footnote{BrandWatch: \url{https://www.brandwatch.com}} to collect data from six major social media platforms, including Reddit, X, Tumblr, TikTok, YouTube, and Facebook. 
We focus on posts published between January 1, 2023, shortly after ChatGPT's public release in late 2022~\cite{Marr2023}, and September 30, 2025, a nearly three-year time frame that captures the rapid evolution of this discourse.
Each search query includes the LLM models and mental health condition (see Table~\ref{tab:search_terms}). 
We only focus on English-language posts. Applying these criteria yields $129,543$ posts. We then remove re-posts and retain only unique texts for result analysis, resulting in a final dataset of $112,698$ posts.

\begin{table}[t]
\centering
\footnotesize
\caption{Search query components for data collection}
\vspace{-2mm}
\label{tab:search_terms}
\renewcommand{\arraystretch}{1.15}
\begin{tabular}{p{1.4cm}p{6.5cm}}
\toprule
\textbf{Field} & \textbf{Search Conditions} \\
\midrule
\textbf{Platform} &
``Reddit'', ``X (formerly Twitter)'', ``Facebook'', ``Instagram'', ``YouTube'', ``TikTok'' \\
\midrule
\textbf{Date Range} & Jan 1, 2023 -- Sep 30, 2025 \\
\midrule
\textbf{Search Terms} &
(\textbf{LLM-chatbot terms}: ``large language model'', ``LLM'', ``GPT'', ``ChatGPT'', ``Claude'', ``Llama'', ``Gemini'', ``Mistral'', ``DeepSeek'', ``Qwen'') \textbf{AND} \\
& (\textbf{Mental health terms}: ``mental health'', ``mental disorder'', ``mental illness'', ``schizophrenia'', ``bipolar disorder'', ``manic depression'', ``cyclothymia'', ``depression'', ``MDD'', ``anxiety disorder'', ``panic disorder'', ``social phobia'', ``OCD'', ``PTSD'', ``eating disorder'', ``anorexia'', ``bulimia'', ``autism spectrum disorder'', ``Asperger's syndrome'', ``autistic'', ``ADHD'', ``attention deficit'', ``attention-deficit hyperactivity disorder'', ``conduct disorder'', ``anger issues'', ``aggressive behavior'', ``personality disorder'') \\
\bottomrule
\end{tabular}
\vspace{-4mm}
\end{table}

\vspace{-1mm}
\subsection{Conceptual Design}

This section explains \emph{how} and \emph{why} we conceptually structure the information extracted. Because social media posts often contain unstructured, ambivalent, or value-laden reflections, we design a schema that identifies both what users report and why those experiences matter, using five labels: \textsc{llm\_product}, \textsc{impact\_sentiment}, \textsc{mental\_health\_condition}, \textsc{user\_perspective}, and \textsc{user\_value}.

To fully capture the reports of users, our schema includes three descriptive labels: \textsc{llm\_chatbot}, identifying which chatbot is discussed; \textsc{impact\_sentiment}, capturing the sentiment or effect users attribute to the interaction; and \textsc{mental\_health\_condition}, identifying the condition referenced in the post. These labels represent the key observable elements in each post---what the user is discussing, how they describe the chatbot’s impact, and which mental health contexts are involved.

The remaining two labels, \textsc{user\_perspective} and \textsc{user\_value}, extend the schema beyond descriptions to capture the experiential and ethical dimensions of each post. We include these dimensions because understanding LLM use for mental health requires examining both \textit{what} users experience and \textit{why} it matters. User perspectives capture the functional aspects of human-LLM interaction (e.g., emotional support \cite{Lawrence2024, song2025typing}), while user values reveal the ethical considerations shaping how those experiences are interpreted (e.g., autonomy when fearing dependency) \cite{friedman2013value}.

\vspace{4px}\noindent$\bullet$~{\bf Inductive analysis of user perspective.} We adopt a bottom-up, inductive approach: Informed by prior literature regarding how users perceive and engage with LLMs for mental health purposes, we compile four possible themes, including \textit{social support}~\cite{Lawrence2024, sharma2024facilitating, vaidyam2019chatbots, song2025typing}, \textit{(over)reliance}~\cite{fang2025ai}, 
\textit{inaccuracies} (e.g., hallucination \cite{jin2025applications, hipgrave2025balancing}), and \textit{ease of use} \cite{lai2017literature}, with an additional \emph{other} category to capture perspectives emerging from users' reports. 

\vspace{4px}\noindent$\bullet$~{\bf Deductive analysis of user value.} We adopt a top-down, deductive approach grounded on the VSD framework~\cite{friedman2013value}. VSD posits that technology is not value-neutral but actively shapes and is shaped by the moral values of its stakeholders. We include $12$ core values from the VSD literature, including \emph{human welfare}, \emph{autonomy}, \emph{privacy}, \emph{informed consent}, \emph{trust}, \emph{accountability}, \emph{fairness}, \emph{intellectual property}, \emph{ownership}, \emph{identity}, \emph{calmness}, and \emph{sustainability}, as a pre-defined codebook for value extraction.

These two dimensions together form a richer analytical framework: the bottom-up perspective categories reflect use cases as they emerge from user discourse, and the top-down value situates these experiences within established ethical AI design principles. This integration enables us to identify not only patterns of LLM use across conditions but also value alignments that inform the design of safer, more ethically responsive LLMs for mental health support.


\vspace{-1mm}
\subsection{Information Extraction}
To extract structured information from the collected social media posts, we design an LLM-assisted annotation pipeline. This pipeline leverages GPT-4.1-mini~\cite{ChatGPT41mini} to identify and categorize relationships between LLM use and mental health as discussed in each post. We aim to capture multiple dimensions of user experiences, enabling a comprehensive analysis of how users perceive and interact with LLM chatbots in mental health contexts.

\vspace{4px}\noindent\textbf{Prompt design.}
Based on our conceptual design, we create a structured prompt that guided LLM to identify the information from each post. The prompt follows a system–task structure: the \emph{system prompt} established the LLM's role as an expert analyst of LLM–mental health relationships, while the \emph{task prompt} specifies the extraction schema and provides detailed guidelines for each dimension. Our prompting approach produces a list of tags grounded in the social media post.

\vspace{4px}\noindent\textbf{Chain-of-Thoughts (CoT) and few-shot prompting.}
\noindent We implement CoT reasoning with the few-shot learning approach~\cite{Wei2023, Brown2020}, where each prompt includes a dedicated \emph{supporting quote}. We define a \emph{supporting quote} as the specific portion of a social media post that provides explicit evidence for determining the final tag. Examples of supporting quotes are listed in Appendix~\ref{tab:mental-health-llm-examples}. Next, we carefully select few-shot examples that illustrate the information extraction procedure. These few-shot examples guide the model to effectively handle diverse cases, including posts with multiple distinct impacts, comparisons across different LLM products, and posts expressing ambivalent perspectives.




\vspace{4px}\noindent\textbf{Pipeline implementation.}
We implement the extraction pipeline using GPT-4.1-mini~\cite{ChatGPT41mini} as the base model, selected for its balance of cost efficiency and extraction performance, as well as its reliability in structured information extraction. Each social media post is analyzed independently, with the structured \textsc{JSON} output parsed and validated programmatically. We set the temperature to 0 to minimize variability and ensure reproducibility. The pipeline processes posts in batches to optimize API usage while maintaining extraction quality. All extracted data are stored in a structured database for subsequent statistical and qualitative content analysis (see sample output in Appendix Table~\ref{tab:mental-health-llm-examples}).

\vspace{-1mm}
\subsection{Performance Validation}
\vspace{4px} \noindent {\bf Procedure.}
Two researchers independently validate the LLM-assisted tagging results across five dimensions on a random sample of 162 annotated instances from 100 social media posts. To prioritize accuracy, we adopt a strict consensus protocol: the annotation inferred by LLM is considered correct only if both human annotators agree with the generated label; otherwise, the sample is flagged as incorrect. In other words, if only one or none of the researchers agree, the human annotation is marked as the opposite of the annotation inferred by the LLM. 

\vspace{4px}\noindent {\bf Validation Results.}
The final inter-rater reliability scores, calculated via the Krippendorff's alpha~\cite{Klaus2011}, demonstrate high agreement across all dimensions. Specifically, we observe high consistency for \textsc{LLM\_Product} ($\alpha=1.00$), \textsc{User\_Perspective} ($\alpha=0.94$), and \textsc{Mental\_Health\_Condition} ($\alpha=0.91$). Substantial agreement is also achieved for the more subjective categories of \textsc{LLM\_Impact} ($\alpha=0.93$) and \textsc{User\_Value} ($\alpha=0.95$). These high reliability coefficients validate the robustness of the LLM-assisted coding process, supporting its application of the model to the full dataset. Following this application, the dataset is reduced from 38,450 to 38,352 entries by removing 98 rows containing missing or incomplete data.

\vspace{-1mm}
\subsection{User Perspective Standardization}



We use a combined manual and LLM-assisted method to analyze \textsc{User\_Perspective}. This process consists of two key steps: One author perform qualitative content analysis~\cite{Hsieh2005ThreeAnalysis:} on a random sample of 100 entries from this column to develop a preliminary codebook. This mapping is collaboratively validated by a second author, with both authors achieving $100$\% agreement rate to establish a final codebook of 12 distinct themes (see Appendix Table~\ref{tab:codebook}). 

We then leverage LLM-as-judge with this codebook to annotate user perspectives extracted from $38,352$ posts. To validate the annotation, a separate sample of 100 rows is manually reviewed, and both authors again achieve $100$\% agreement on the annotation accuracy after discussion. In the final data refinement phase, a total of $75$ rows are excluded due to ``N/A'' outputs or invalid value entries. This resulted in a final analytical dataset of $38,277$ posts.

\section{Results}

\subsection{Sentiment across mental conditions in LLM}
\label{RQ1}
\begin{figure*}[t]
\centering
\includegraphics[width=\textwidth]{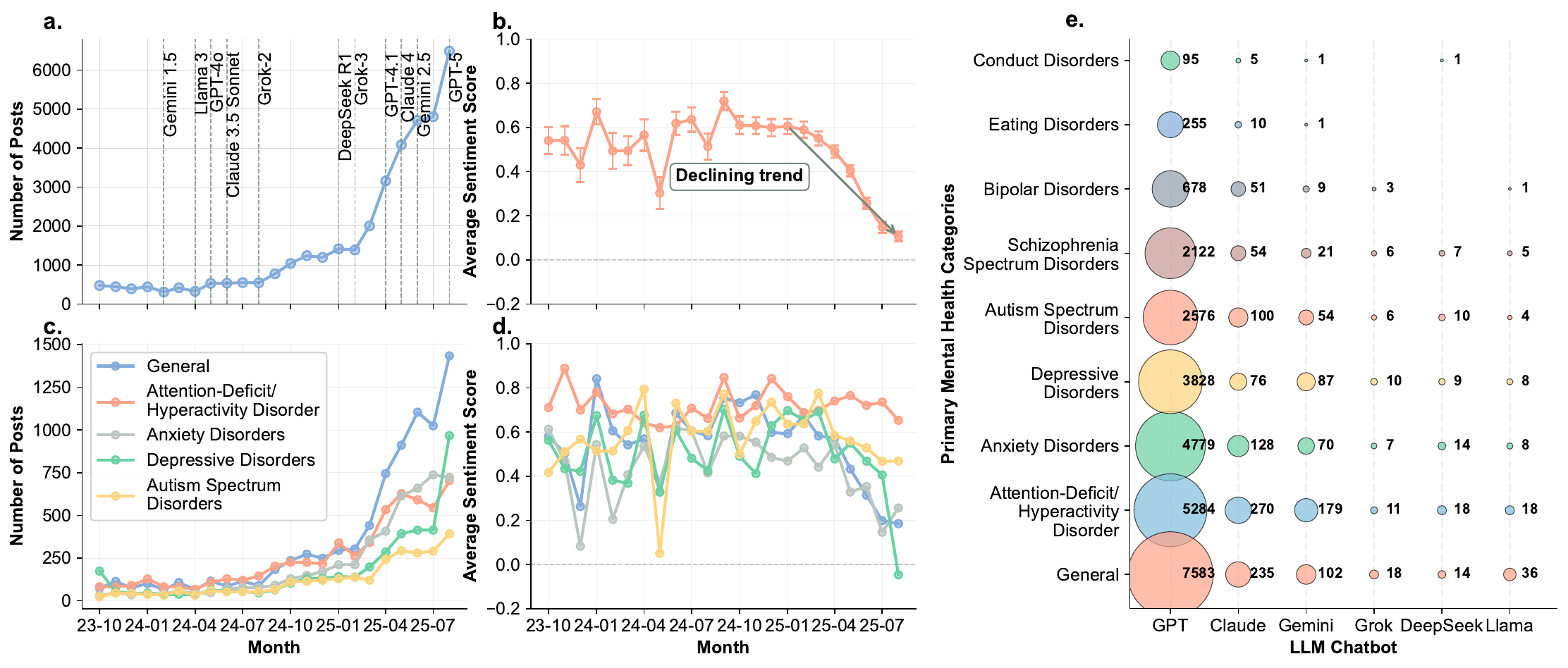}
\vspace{-0.3in}
\caption{Time-series trends:
(a) Monthly post volume,
(b) Average sentiment over time,
(c) Post volume by mental health condition,
(d) Average sentiment by mental health condition, and
(e) Fingerprint of LLM chatbot across mental health conditions.}
\vspace{-0.1in}
\label{fig:timeseries}
\end{figure*}

\begin{figure*}[t]
  \centering
  \includegraphics[width=0.92\textwidth]{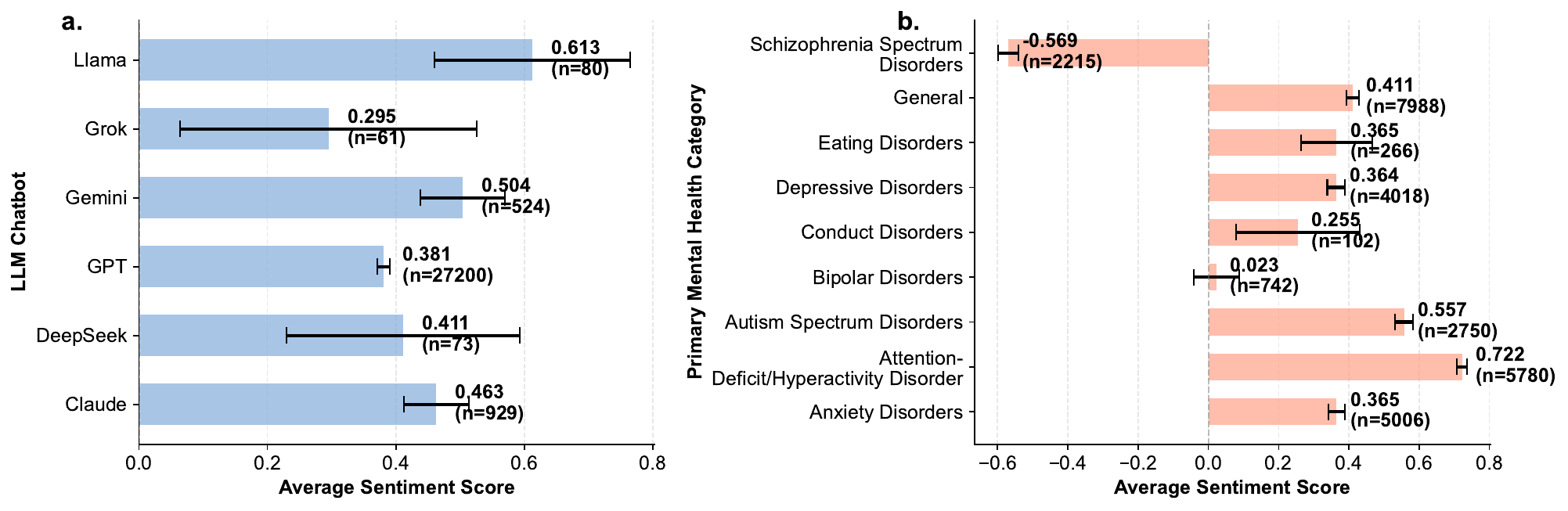}
  \vspace{-2mm}
  \caption{Average sentiment variation across: (a)  LLM chatbots. (b) Mental health conditions.}
  \label{fig:sentiment}
\end{figure*}

To address RQ1, we integrate three complementary analyses: (1) a time-series analysis (Figure~\ref{fig:timeseries}a - d) tracking how discussions linking LLM chatbots and mental health change over time; (2) a cross-condition and chatbot ``fingerprint'' (Figure~\ref{fig:timeseries}e) mapping where each chatbot appears across mental health conditions; and (3) a sentiment analysis (Figure~\ref{fig:sentiment}) measuring emotional tones across chatbots and conditions. These analyses identify when conversations scale, which conditions are associated with which chatbots, and how users interpret interactions with LLMs.

\vspace{4px}\noindent\textbf{A. Public engagement with LLMs for mental health surges but sentiments decline.}
Public discussion around LLM chatbots and mental health increases more than fivefold from early 2024 to mid-2025 (Figure~\ref{fig:timeseries}a). This surge aligns closely with major model releases --- Claude 3.5 Sonnet, GPT-4.1, Gemini 2.5, and GPT-5 --- suggesting that advances in LLM capabilities coincide with rising public interest in using them as mental health tools.

Condition-specific trends further show that discussions related to ADHD, ASD, anxiety disorders, and depressive disorders grow the fastest, with ADHD and ASD accelerating most sharply by mid-2025 (Figure~\ref{fig:timeseries}c). These increases suggest that users are turning to LLMs for a wider set of day-to-day needs, including emotional expression as well as structured assistance such as organizing thoughts, managing routines, or navigating overwhelming situations.

Despite this increasing engagement, the average sentiment displays a steady downward trend beginning in early 2025 (Figure~\ref{fig:timeseries}b). As LLM use has broadened into more diverse contexts, public discussions become increasingly ambivalent, potentially reflecting heightened awareness of risks such as over-reliance, misinformation, or inappropriate emotional reinforcement. Sentiment also varies substantially across mental health conditions (Figure~\ref{fig:timeseries}d), revealing that user experiences are far from uniform---ranging from relief and usefulness to frustration, caution, or distrust.

\vspace{4px}\noindent\textbf{B. GPT dominates overall volume, but mental health conditions show distinct patterns.}
The fingerprint (Figure~\ref{fig:timeseries}e) reveals how LLMs are discussed across mental health conditions. GPT dominates the discussion; for example, it appears in $7,583$ ADHD posts, $3,472$ ASD posts, and $2,122$ depression posts, reflecting its accessibility and widespread adoption. This makes GPT the default point of reference when people discuss LLMs in mental health contexts. A closer look shows distinct preference patterns. Claude appears proportionally more often in ADHD- and ASD-related posts, while Gemini is mentioned more frequently in general and anxiety-related discussions. In contrast, open-source models such as Llama, DeepSeek, and Grok appear infrequently and mainly within ADHD and ASD posts.

In addition, mental health conditions differ substantially in how often they appear in LLM-related discussions. General mental health posts, ADHD, ASD, anxiety disorders, and depressive disorders form the largest clusters, as they involve daily challenges suited to informational or emotional support. Conditions involving elevated clinical risk, such as schizophrenia spectrum disorders, bipolar disorder, eating disorders, and conduct disorders, appear less frequently. Their lower volumes suggest more cautious or sporadic engagement, yet their inclusion indicates that LLM-related discussions have expanded into high-risk mental health contexts.


\vspace{4px}\noindent\textbf{C. Sentiment varies across LLMs and mental health conditions.}
Figure~\ref{fig:sentiment} shows differences in sentiment across LLM chatbots. Llama and Gemini receive the highest average sentiment scores ($0.613$ and $0.504$), while GPT, despite dominating overall volume, shows a more moderate score ($0.381$). These differences indicate that sentiment does not simply follow adoption patterns; rather, it reflects how users evaluate each model's style, responsiveness, and perceive suitability for mental health–related conversations.

Sentiment also differs substantially across mental health conditions. ADHD and ASD discussions show notably positive sentiment (0.722 and 0.557), with users often describing LLMs as helpful for structuring thoughts, managing overload, or providing steady support. Anxiety and depressive disorder posts cluster around moderate sentiment (0.36), indicating mixed or neutral experiences. In contrast, schizophrenia-spectrum discussions show strongly negative sentiment (–0.569), frequently describing confusion, fear, or cases where chatbots' responses appear to intensify or validate delusional thinking. These patterns suggest that users experiencing psychosis-related symptoms may be particularly vulnerable to unintended reinforcement, underscoring the need for condition-aware, context-sensitive design in high-risk mental health settings.

\subsection{Perspectives and Values for Mental Health}

\subsubsection{Value-Perspective Co-Occurrences in User Sentiment}
\label{RQ2.1}
To examine how users' values and perspectives toward LLM-based mental health interactions differ by sentiment, we compute Pointwise Mutual Information (PMI) for each Value–Perspective pair separately for positive and negative posts. PMI quantifies how strongly two codes co-occur beyond what could be expected by chance, after accounting for each code's base rate~\cite{church1990word}. This normalization is crucial because certain values (e.g., Human Welfare) and perspectives (e.g., Instrumental Support) appear far more often than others, and raw co-occurrence counts could obscure meaningful associations. For a Value $v$ and Perspective $p$, the PMI is defined as:


\vspace{-5mm}
\begin{align}
    \text{PMI}(v,p) = \log \bigl( P(v,p) \, / \, (P(v)\,P(p)) \bigr)
\end{align}

where $P(v,p)$ is the probability that the two codes co-occur, $P(v)$ is the probability of the Value code appearing, and $P(p)$ is the probability of the Perspective code appearing. To examine sentiment-specific associations, we compute a PMI difference metric:

\vspace{-5mm}
\begin{align}
\Delta \text{PMI}(v,p)
= \text{PMI}(v,p \mid \text{positive})
- \text{PMI}(v,p \mid \text{negative})
\end{align}
\vspace{-5mm}

\begin{figure}[t]
\centering
\includegraphics[width=\linewidth]{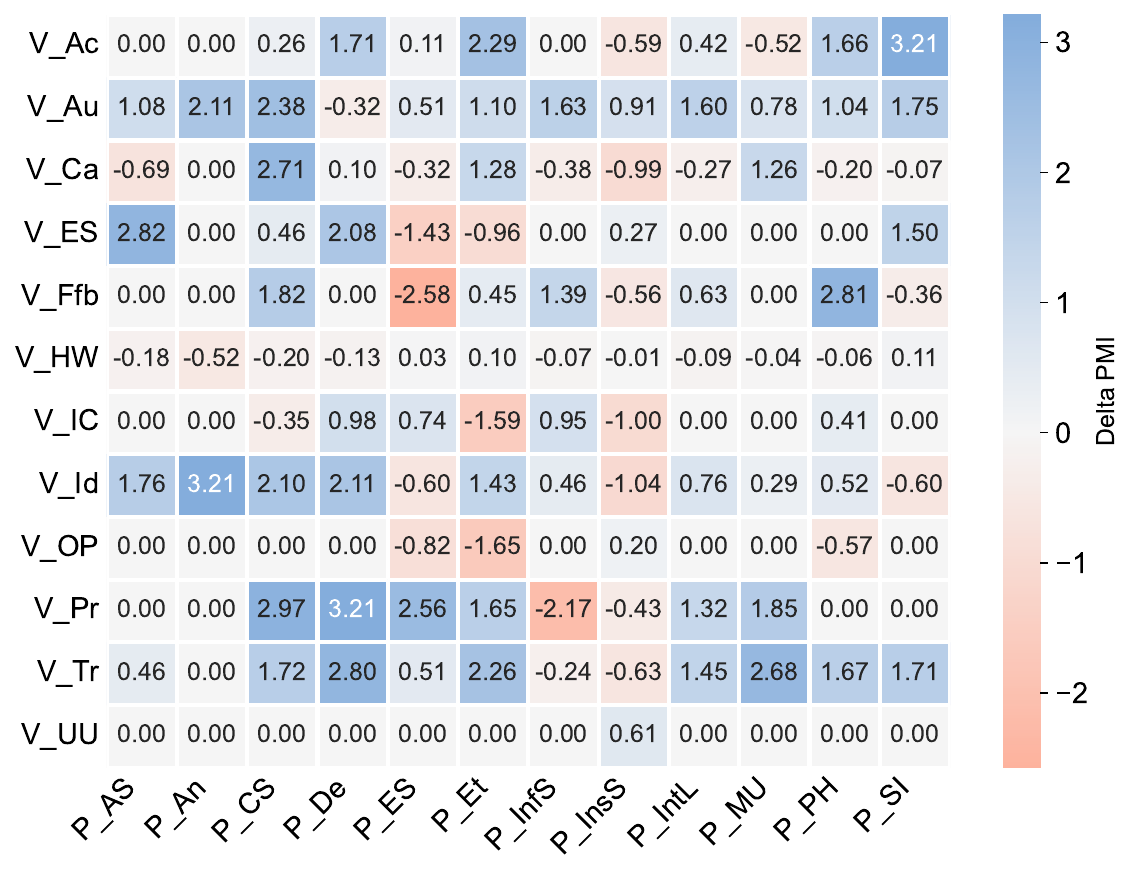}
\vspace{-8mm}
\caption{Heatmap on Pointwise Mutual Information (PMI) difference for Value‑Perspective co‑occurrences across sentiment groups.
\footnotesize{\textnormal{
(For abbreviations, see Figure~\ref{fig:heatmap} caption.)}}}
\vspace{-5mm}
\label{fig:heatmap_pmi}
\end{figure}

A positive $\Delta \text{PMI}$ indicates a stronger association in posts with \textit{positive} sentiment, whereas a negative value reflects a stronger association in \textit{negative} sentiment posts. The magnitude of $\Delta \text{PMI}$ represents the extent to which the Value–Perspective relationship diverges across sentiment groups. The resulting heatmap \textbf{(Figure~\ref{fig:heatmap_pmi})} highlights clusters where these associations diverge, with blue cells indicating stronger links in positive posts and red cells indicating stronger links in negative posts.

\vspace{4px}\noindent\textbf{A. Positive sentiment: Identity alignment, accountability, and safe reliance.} 
Several Value–Perspective pairs show stronger associations in \emph{positive} sentiment posts. The largest effect is observed for \textit{Identity $\times$ Anthropomorphism} (\(\Delta \text{PMI} = 3.21\)), indicating that users with positive experiences often describe LLMs as human-like or personally relatable, sometimes framing them as companions or identity-affirming agents. A similarly strong positive shift appears for \textit{Accountability $\times$ Sociocultural Impact} (\(\Delta \text{PMI} = 3.21\)), suggesting that positive sentiment often emphasizes LLM alignment with sociocultural expectations and responsible behavior.

Other notable positive co-occurrences include \textit{Calmness $\times$ Clinical Skepticism} (\(\Delta \text{PMI} = 2.71\)) and \textit{Environmental Sustainability $\times$ Appraisal Support} (\(\Delta \text{PMI} = 2.82\)), indicating that users sometimes view LLM-assisted reflection as both reassuring and consistent with ecological or longer-term well-being values. 

Privacy-related values show clear positive associations: 
\textit{Privacy $\times$ Clinical Skepticism} (\(\Delta \text{PMI} = 2.97\)), \textit{Privacy $\times$ Dependency} (\(\Delta \text{PMI} = 3.21\)), and \textit{Privacy $\times$ Emotional Support} (\(\Delta \text{PMI} = 2.56\)). These results suggest that users with positive sentiment often perceive LLMs as private, low-risk environments in which emotional disclosure, reliance, and reflective thinking can more comfortably occur. Finally, \textit{Trust}-related pairings, including \textit{Trust $\times$ Dependency} (\(\Delta \text{PMI} = 2.80\)) and \textit{Trust $\times$ Maladaptive Usage} (\(\Delta \text{PMI} = 2.68\)), indicate that trust in LLMs may coincide with increased reliance.

\vspace{4px}\noindent\textbf{B. Negative sentiment: Biases, risks, and unmet needs.}
A small set of co-occurrences shows strong associations in \emph{negative} sentiment posts, but these highlight concentrated areas of concern. The strongest negative association is \textit{Freedom from Bias $\times$ Emotional Support} (\(\Delta \text{PMI} = -2.58\)), suggesting that negative evaluations often stem from perceptions that LLMs provide biased or inadequate emotional responses. Meanwhile, \textit{Freedom From Bias $\times$ Psychological Harm} (\(\Delta \text{PMI} = 2.81\)) shows that discussions of bias are also tied to perceived risks of misinformation or emotional harm.

Two additional pairings, including \textit{Informed Consent $\times$ Ethics} (\(\Delta \text{PMI} = -1.59\)) and \textit{Ownership and Property $\times$ Ethics} (\(\Delta \text{PMI} = -1.65\)), highlight that negatively valenced posts often framed LLM interactions around ethical concerns regarding data usage, consent, and ownership. A strong negative association emerges for \textit{Privacy $\times$ Informational Support} (\(\Delta \text{PMI} = -2.17\)), suggesting that when users express negative sentiment, privacy concerns are linked to unreliable or insufficient informational assistance.

Overall, positive-sentiment discourse is characterized by strong connections between values such as identity, privacy, accountability, and trust, and perspectives involving emotional support, anthropomorphism, dependency, and sociocultural alignment. In contrast, negative-sentiment posts concentrated on a narrower but sharper cluster of concerns, linking values such as freedom from bias, informed consent, and ownership with perspectives related to emotional inadequacy, unclear ethical boundaries, and psychological risk. These patterns reveal two qualitatively distinct relational structures in how users make sense of LLMs in mental-health contexts: one emphasizing support, alignment, and reassurance, and the other emphasizing risk, fairness, and unmet emotional needs.

\subsubsection{Mental Health Condition-Specific Sentiment, Values, Perspective}
\label{RQ2.2}
To analyze the relationships between mental health conditions and user sentiment, values, and perspectives on LLM chatbot use, we employ Pearson's Chi-square tests of independence \cite{pearson1900x} to evaluate statistical significance with an alpha level of $\alpha = .05$. Effect sizes are quantified using Cramér's V \cite{akoglu2018user}: values $< 0.05$ denote a weak effect, 0.1-0.15 a moderate effect, and $> 0.15$ a strong effect. We further decompose global dependencies using adjusted standardized residuals \cite{haberman1973analysis}. Following Agresti \cite{agresti2010analysis}, residuals exceeding $\pm 2.0$ are identified as significant deviations from the null hypothesis of independence, where positive values ($> 2.0$) indicate a strong association and negative values ($< -2.0$) indicate a significant dissociation. We conduct all analyses in Python.

Pearson's Chi-square tests confirmed that mental health conditions are strongly associated across user values, perspectives, and sentiments \ref{tab:chisquare_results}). Specifically, mental health conditions exhibit a stronger association with both sentiment (Cramér's $V = 0.27$) and perspective ($V = 0.18$), suggesting that a user's diagnosis or self-reported conditions shape the emotional tone and framing of their narratives on social media. The association with value, although comparatively subordinate, has a moderate effect ($V = 0.09$). 

\begin{table}[t]
    \centering
    \footnotesize
    \caption{Pearson’s Chi-square tests of independence.}
    \vspace{-4mm}
    \label{tab:chisquare_results}
    \begin{tabularx}{\columnwidth}{@{} >{\raggedright\arraybackslash}X c c c l @{}}
        \toprule
        \textbf{Comparison} & \textbf{Chi-Square ($\chi^2$)} & \textbf{\textit{p}-value} & \textbf{Cramér's $V$} & \textbf{Effect Size} \\ 
        \midrule
        Mental Health vs. Sentiment   & 4,646.28  & $< .001$ & 0.27 & Strong \\
        Mental Health vs. Perspective & 10,425.19 & $< .001$ & 0.18 & Strong \\
        Mental Health vs. Value       & 2,561.30  & $< .001$ & 0.09 & Moderate \\ 
        \bottomrule
        \multicolumn{5}{@{}p{\columnwidth}@{}}{\textit{Note:} Effect sizes are interpreted based on Akoglu (2018) \cite{akoglu2018user}, where $V > 0.15$ indicates a strong effect.}
    \end{tabularx}
    \vspace{-3mm}
\end{table}

\begin{figure*}[t]
\centering
\includegraphics[width=\textwidth]{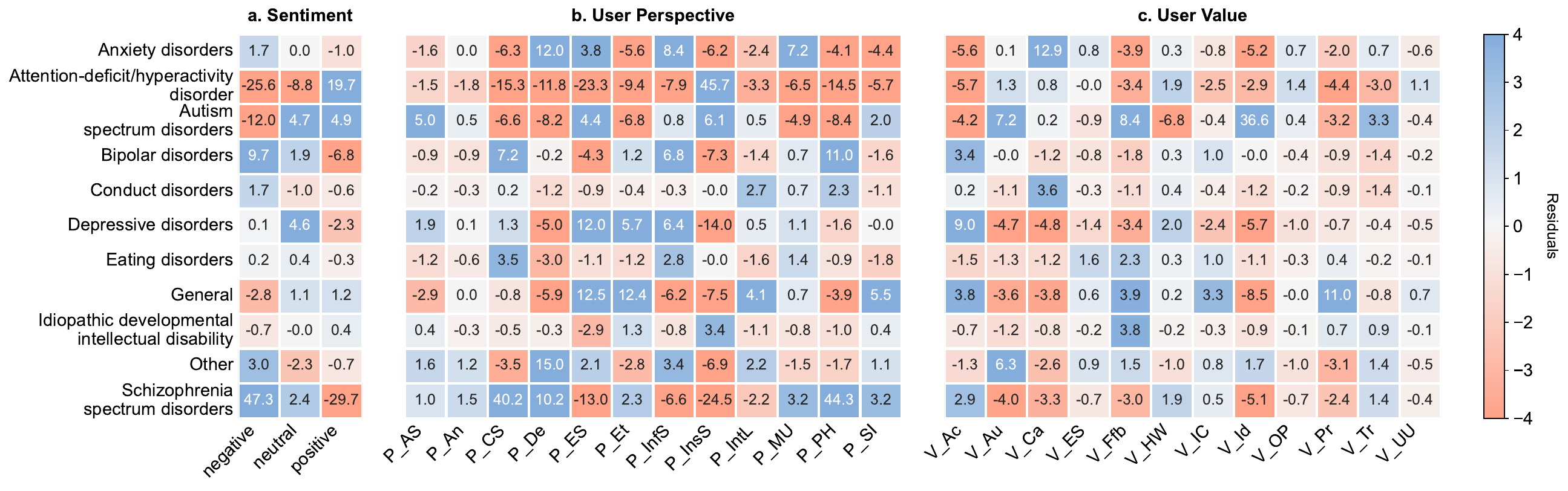}
\vspace{-8mm}
\caption{Heatmap on the strength of associations between mental health (MH) conditions and user interaction variables based on Adjusted Standardized Residuals (ASR). 
(a) ASR of MH Condition vs. Sentiment. 
(b) ASR of MH Condition vs. Perspective. 
(c) ASR of MH Condition vs. Value. 
\footnotesize{\textnormal{Abbreviations: 
\textit{Values (V):} 
V\_Ac: Accountability; 
V\_Au: Autonomy; 
V\_Ca: Calmness; 
V\_ES: Environmental Sustainability; 
V\_Ffb: Freedom From Bias; 
V\_HW: Human Welfare; 
V\_IC: Informed Consent; 
V\_Id: Identity; 
V\_OP: Ownership and Property; 
V\_Pr: Privacy; 
V\_Tr: Trust; 
V\_UU: Universal Usability. 
\textit{Perspectives (P):} 
P\_An: Anthropomorphism; 
P\_AS: Appraisal Support; 
P\_CS: Clinical Skepticism; 
P\_De: Dependency; 
P\_ES: Emotional Support; 
P\_Et: Ethics; 
P\_InfS: Informational Support; 
P\_InsS: Instrumental Support; 
P\_IntL: Interaction Limitations; 
P\_MU: Maladaptive Usage; 
P\_PH: Psychological Harm; 
P\_SI: Sociocultural Impact.}}}
\label{fig:heatmap}
\vspace{-2mm}
\end{figure*}

\vspace{4px}\noindent\textbf{A. Association between mental health conditions and sentiment.} 
(1) Neurodivergent conditions align with positive sentiment (Figure \ref{fig:heatmap}b). User narratives describing ADHD appeared most often in positive sentiment ($ASR = 19.7$), followed by autism spectrum disorders ($4.9$). These users consistently describe their interactions with LLM chatbots in favorable terms. The absence of negative sentiment for ADHD ($-25.6$) and a dissociation from emotional support ($P\_ES$, $-23.3$) and dependency ($P\_De$, $-11.8$) reinforce the utility of these interactions.
(2) Severe psychiatric disorders are characterized by negative sentiment. Schizophrenia spectrum disorders are overwhelmingly linked to negative sentiment ($47.3$) and dissociated from positive ones ($-29.7$). Bipolar disorders display a similar but less extreme pattern. 
(3) Depressive disorders are associated with neutral sentiment. Depressive disorders show an isolated association with neutral sentiment ($4.6$), with negligible associations in positive ($-2.3$) or negative ($0.1$) sentiments.

\vspace{4px}\noindent\textbf{B. Association between mental health conditions and user perspectives.}
(1) Users mentioning ADHD often frame their interactions with LLM chatbots through instrumental support (Figure \ref{fig:heatmap}c for $P\_InsS$, $45.7$). The strong dissociation from clinical skepticism ($-15.3$) further indicates that these users prioritize utilizing the LLM chatbot's capabilities rather than questioning its validity.
(2) Schizophrenia spectrum disorders appear most frequently in the clinical skepticism ($P\_CS$, $40.2$) and psychological harm ($P\_PH$, $44.3$). This suggests that the user's intent is not to seek assistance from LLM chatbots, but to test the LLM chatbot's medical knowledge boundaries or navigate potential safety failures. The dissociation from instrumental support ($-24.5$) indicates a lack of trust or inability to utilize the chatbot for daily tasks.
(3) Users mentioning autism spectrum disorders are associated with instrumental support ($P\_InsS$, $6.1$) and appraisal support ($P\_AS$, $5.0$). This suggests that users employ the LLM chatbot not only for tangible tasks but also to obtain constructive feedback.
(4) Users mentioning depressive disorders stress emotional and informational support while dissociating from instrumental support. User narratives regarding depressive disorders show a strong positive association with emotional support ($P\_ES$, $12.0$) and informational support ($P\_InfS$, $6.4$), while exhibiting the strongest negative dissociation from instrumental support ($P\_InsS$, $-14.0$). 
(5) Anxiety disorders are strongly associated with dependency ($P\_De$, $12.0$) and maladaptive usage ($P\_MU$, $7.2$). This suggests that user narratives frequently describe a heightened reliance on the LLM chatbots.
(6) Similar to schizophrenia spectrum disorders, bipolar disorders are associated with psychological harm ($P\_PH$, $11.0$) and clinical skepticism ($P\_CS$, $7.2$). However, the dissociation from instrumental support is less pronounced for bipolar disorders ($-7.3$) compared to schizophrenia ($-24.5$). This implies that user narratives regarding Bipolar disorders do not describe the same degree of rejection of the LLM chatbot's utility.

\vspace{4px}\noindent\textbf{C. Association between mental health conditions and user values.}
(1) User narratives regarding autism spectrum disorders exhibit a focus on identity (see Figure \ref{fig:heatmap}d). These users show the highest value association, linking strongly to identity ($V\_Id$, $36.6$), freedom from bias ($V\_Ffb$, $8.4$), and Autonomy ($V\_Au$, $7.2$). 
(2) Users who mention anxiety emphasize calmness value. Anxiety disorders appear most often in the calmness ($V\_Ca$) ($12.9$), with a negative association with accountability value ($V\_Ac$, $-5.6$). 
(3) Users mentioning depressive disorders stress accountability while dissociating from identity. Depressive disorders discussions show a significant association with accountability ($V\_Ac$, $9.0$). However, they exhibit a strong dissociation from identity ($V\_Id$, $-5.7$). 
(4) ADHD narratives show a broad dissociation. Narratives regarding ADHD display negative associations across almost all value categories, notably accountability ($V\_Ac$, $-5.7$) and privacy ($V\_Pr$, $-4.4$), and show minimal residuals across other values.
(5) Schizophrenia spectrum disorder discussions are dissociated from identity and autonomy. Similar to depressive disorders, schizophrenia spectrum narratives showed a strong dissociation from identity ($V\_Id$, $-5.1$) and autonomy ($V\_Au$, $-4.0$).

\section{Discussion}

\subsection{Condition-Specific Patterns in LLM Usage}

As users increasingly turn to LLM chatbots for mental health support, our analysis reveals significant disconnects between LLM chatbot design and the user needs. Our time-series analysis (Section~\ref{RQ1}) reveals that current LLMs frequently fail to meet the nuanced needs from various mental health conditions. The variation in user sentiment, from highly positive in neurodivergent discussions to strongly negative in psychosis-related ones, further demonstrates that user experiences are far from uniform (Figure~\ref{fig:timeseries}d).


\textbf{Instrumental support for neurodivergence.} Narratives from users with ADHD and ASD reveal that LLM chatbot interactions center on functional utility rather than emotional connection. Rather than describing LLMs as therapists, these posts frequently detail how users leverage LLM chatbots to scaffold executive functioning and manage daily routines \cite{bucher2025s}. Furthermore, the strong association between ASD and \textit{Identity} suggests that positive outcomes stem from LLMs aligning with their self-conception and autonomy.


\textbf{Concerns on LLM for psychosis spectrum disorders.} User discussions related to schizophrenia spectrum show the strongest negative sentiment and are significantly associated with \textit{Clinical Skepticism} and \textit{Psychological Harm}. These users report encounters with confusing, inaccurate, or clinically unsafe responses. Our findings support prior work on LLM hallucinations  \cite{clegg2025shoggoths, Gold2025} that LLMs' generative variability could be a safety risk in high-stakes contexts.


\textbf{Emotion regulation and over-reliance in LLM use.} Posts regarding depressive and anxiety disorders emphasize \textit{Emotional Support} and \textit{Informational Support} (Figure \ref{fig:heatmap}c), often describing LLMs as a non-judgmental space for cognitive restructuring or venting \cite{sharma2024facilitating, feng2025effectiveness}. However, anxiety-related discussions also highlight the associated risks including \textit{Dependency} and \textit{Maladaptive Usage}. While the LLM chatbots may provide immediate calming effects, these narratives suggest that for users with anxiety, the accessibility of LLM chatbots can fuel cycles of compulsive reassurance-seeking \cite{fang2025ai}. This suggests that LLM chatbots could risk reinforcing avoidance behaviors in those managing anxiety. Thus, mental health support should not treated as a monolithic umbrella in LLM chatbot interactions but as differentiated pathways, which requires the value-sensitive design implications in Section~5.2.

\subsection{LLM VSD across Mental Health Conditions}
While prior work has documented the benefits of AI chatbots for mental health \cite{vaidyam2019chatbots,feng2025effectiveness}, our findings reveal that user experiences with LLM chatbots reflect different value-perspectives for mental health conditions, thereby exposing the limits of one-size-fits-all safety approaches~\cite{liscio2025value,sadek2024guidelines} that assume all users may seek similar kinds of support. In this sense, we define \textit{\textbf{value-sensitivity}} as an LLM chatbot's capacity to recognize, adapt to, and align with the human values (e.g., autonomy, identity) preferred by different user groups within specific contexts of use. It requires designers to acknowledge that values are not static individual traits, but rather dynamic and situational, shaping users' cognitive states, emotional needs, and sociocultural conditions. Achieving this requires:

\textbf{Operationalizing autonomy for neurodivergence.} Neurodivergent user (e.g., ADHD and ASD)'s emphasis on \textit{Instrumental Support} and \textit{Identity}, reinforced by the link between identity and anthropomorphism (Section \ref{RQ2.1}), suggests LLMs serve primarily as cognitive peers rather than emotional companions \cite{pergantis2025ai, jamshed2025rethinking, namvarpour2025understanding, yankouskaya2025can}. Future designs should shift from generic conversational empathy toward instrumental scaffolding that enables users to customize the structure and tone of LLM interactions (Sections \ref{RQ1} and \ref{RQ2.2}), preserving their autonomy and authentic identity expression.

\textbf{Context-aware safety for psychosis risk. }For psychosis-risk conditions, such as schizophrenia spectrum disorders, strong negative sentiment and associations with \textit{Clinical Skepticism} and \textit{Psychological Harm} suggest the limitations of general-purpose LLMs in high-stakes contexts \cite{poenaruai, preda2025special}. The dissociation from \textit{Instrumental Support} highlights a gap in LLM chatbots' utility and safety, requiring condition-aware filtering mechanisms to detect misinterpretations of delusional language \cite{clegg2025shoggoths}, ensuring clinical safety.

\textbf{Balancing validation and dependency in mood disorders.} Users expressing depression through interacting with LLM chatbots emphasize \textit{Emotional Support} and \textit{Accountability} while dissociating from \textit{Identity}, seeking validation and scaffolding, a demand that extends beyond the generic reassurance typical of LLM interactions \cite{yoo2025ai, sabour2023chatbot}. Meanwhile, anxiety disorders’ association with \textit{Dependency} and \textit{Calmness} highlights tensions between temporary relief and long-term coping \cite{hua2025charting}. Future LLM chatbot designs must strike a balance between support and strategies that promote effectiveness and prevent maladaptive dependency.

\vspace{-2mm}
\subsection{Limitations and Future Directions}
Our study has several limitations that inform future work. First, although our data covers a few major social media platforms, it does not capture the full spectrum of user experiences. Future work should incorporate more social media platforms or conduct interview- or survey-based research to capture perspectives from users who do not publicly discuss LLM use and mental health. Second, our analysis relies on self-reported posts without access to clinical health data such as electronic health records. Integrating such data in privacy-preserving ways could allow for stronger validation of risk patterns and health outcomes associated with LLM use in future work. Finally, despite our rigorous LLM-assisted extraction pipeline, our data analysis remains constrained by model accuracy. Improving these pipelines through enhanced prompt design and adversarial testing in future work could further strengthen the reliability of analyses.

 \section{Conclusions}

This study provides a large-scale characterization of how people use and evaluate LLM chatbots in mental health contexts. Using crowdsourced social media data and an LLM-assisted pipeline grounded in VSD, we identify condition-specific patterns: neurodivergent conditions such as ADHD and ASD show predominantly positive engagement, whereas higher-risk disorders, including the schizophrenia spectrum, exhibit negative sentiment and concerns about clinical accuracy and psychological harm. Users’ perspectives and values further reveal differing expectations around autonomy, identity, privacy, and emotional safety. Overall, engagement with LLMs is generally positive, although sentiment tends to decline over time. Their impact varies by condition and value priorities, underscoring the need for condition-specific, value-sensitive LLM systems for mental health support.



\bibliographystyle{ACM-Reference-Format}
\bibliography{main}

\appendix
\section{Appendix}

\subsection{Data Ethics Statement}

Our study analyzes publicly available social media posts related to LLM use in mental health contexts and is conducted with careful attention to data ethics. The university's Institutional Review Board (IRB) determined the project to be exempt under the category of secondary research using publicly available, pseudo-anonymous data. We recognize ongoing ethical debates surrounding the use of public online content, especially when posts involve sensitive health disclosures, and therefore adopt a conservative approach to privacy and identifiability. Prior to analysis, all datasets are fully anonymized by removing usernames, metadata, and other identifying information. Data are stored on secure, access-controlled systems available only to the research team. We further emphasize that our goal is not to evaluate individual users but to characterize aggregate patterns in public discourse, consistent with best practices for ethically responsible use of web data for social good.

\subsection{Thematic Codebook of User Perspectives}
Table~\ref{tab:codebook} summarizes the full set of thematic codes used to annotate user perspectives in our analysis. It provides the definitions, example initial codes, and frequency distributions for each perspective category, offering comprehensive user-reported experiences in our dataset through qualitative content analysis.

\begin{table*}[t!]
\centering
\scriptsize
\caption{Thematic codebook of user perspectives on LLM chatbots.}
\vspace{-3mm}
\label{tab:codebook}
\resizebox{0.9\textwidth}{!}{
\begin{tabularx}{0.9\textwidth}{@{} p{2.4cm} >{\raggedright\arraybackslash}X >{\raggedright\arraybackslash}X r @{}}
\toprule
\textbf{Theme} & \textbf{Definition} & \textbf{Example initial codes from LLM-assisted annotation} & \textbf{Frequency (n, \%)} \\ \midrule

\textbf{Instrumental Support} (P\_InsS) & 
Use of LLM chatbots for practical assistance, task completion, productivity enhancement, or overcoming functional barriers (e.g., writing, scheduling, formatting). & 
\textit{using LLM for text editing to communicate mental health struggles, efficient accessibility, using LLM to organize thoughts and plan mental health approach} & 
13,071 (34.15\%) \\ 

\textbf{Emotional Support} (P\_ES) & 
Interactions with LLM chatbots driven by a need for empathy, validation, companionship, comfort, or alleviation of loneliness. & 
\textit{seeking comfort and reassurance, emotional support, finding relatability with LLM, using LLM for light entertainment and bonding} & 
9,797 (25.60\%) \\

\textbf{Informational Support} (P\_InfS) & 
Use of LLM chatbots to gather facts, define terms, access educational resources, or obtain objective advice regarding mental health topics. & 
\textit{seeking informational support about mental health and neurochemistry, clarification of LLM role, inquiry about mental health relation to LLM products} & 
5,028 (13.14\%) \\ 

\textbf{Clinical Skepticism} (P\_CS) & 
Critical perspectives regarding the safety, effectiveness, and medical accuracy of LLM chatbots, including concerns about misdiagnosis, crisis failures, and lack of professional oversight. & 
\textit{misdiagnosis, skepticism about effectiveness, failure to respond to crisis, doubt about causation of behavior by LLM, providing unsolicited mental health labeling} & 
3,022 (7.90\%) \\ 

\textbf{Psychological Harm} (P\_PH) & 
Direct negative impacts on the user's mental state caused by LLM chatbots, including the triggering of symptoms, induction of distress, or worsening of conditions. & 
\textit{concern about mental health impact, exacerbation of delusional thoughts, triggering negative mental health episodes, paranoia and conspiracy beliefs} & 
1,734 (4.53\%) \\ 

\textbf{Dependency} (P\_De) & 
Issues related to over-reliance, addiction, difficulty disengaging, or the displacement of necessary human relationships with LLM chatbot interaction. & 
\textit{over-reliance, addiction, withdrawal symptoms, emotional attachment leading to potential over-reliance, fear of losing AI companion} & 
1,714 (4.48\%) \\ 

\textbf{Ethics} (P\_Et) & 
Issues regarding privacy, data governance, content moderation, algorithmic bias, and the moral implications of deploying LLM chatbots for mental health. & 
\textit{privacy concern, accountability, mental health safety update, concern about bias and programming influences, ethical considerations in AI testing} & 
1,234 (3.22\%) \\ 

\textbf{Interaction Limitations} (P\_IntL) & 
Frustrations with LLM chatbots' artificial nature, lack of genuine human connection, repetitive responses, memory failures, or inability to understand nuance. & 
\textit{lack human connection, frustration with LLM behavior, perception of mechanical responses, robotic tone} & 
1,005 (2.63\%) \\ 

\textbf{Sociocultural Impact} (P\_SI) & 
Observations on how LLM chatbots affect culture, the economy, and the mental health profession, including societal judgment and stigma surrounding their use. & 
\textit{stigmatizing language, fear of being perceived as using LLM, concern about societal impact, displacement of therapists, accusation of faking mental disorder} & 
544 (1.42\%) \\ 

\textbf{Appraisal Support} (P\_AS) & 
Use of LLM chatbots for self-reflection, decision-making, interpreting personal behaviors, or gaining perspective on one's identity and mental state. & 
\textit{self-reflection and identity exploration, using LLM to analyze interpersonal interactions, seeking appraisal or diagnostic opinion from LLM} & 
534 (1.40\%) \\ 

\textbf{Maladaptive Usage} (P\_MU) & 
Using LLM chatbots in ways that enable unhealthy behaviors, avoid professional care, or reinforce compulsions (e.g., OCD checking, fueling delusions). & 
\textit{misuse, using LLM to fuel late-night hyperactive projects, reassurance seeking as compulsive behavior, using LLM to verify delusions} & 
503 (1.31\%) \\ 

\textbf{Anthropomorphism} (P\_An) & 
Attributing human traits, consciousness, emotions, or mental health conditions to LLM chatbots (e.g., claiming the bot is depressed or neurodivergent). & 
\textit{metaphorical description of LLM behavior, personification of LLM mental health, attributing ADHD-like traits to LLM, treating AI as sentient} & 
91 (0.24\%) \\

\bottomrule
\end{tabularx}
}
\end{table*}

\subsection{Examples of Information Extraction}

Table~\ref{tab:mental-health-llm-examples} presents representative examples from our LLM-assisted extraction pipeline. For each original social media post, the table includes the full set of extracted fields—LLM chatbot, mental health condition, impact sentiment, user perspective, and user value—derived from the Value-Sensitive Design (VSD) framework. All information is extracted using the GPT-4.1-mini model with temperature set to 0.

\begin{table*}[t]
\centering
\footnotesize
\caption{Representative examples output by the pipeline implementation.}
\vspace{-1mm}
\label{tab:mental-health-llm-examples}
\renewcommand{\arraystretch}{1.25}
\resizebox{0.9\textwidth}{!}{
\begin{tabularx}{\textwidth}{@{}X X@{}}
\toprule
\textbf{Original post} & \textbf{LLM-extracted output} \\
\midrule


\begin{minipage}[t]{\linewidth}
I have been citing from the Involuntary Psychiatric Treatment Act (IPTA), the Mental Health Patient Rights documents, and the Canadian Charter of Rights and Freedoms. I did make one mistake when I referred to section 14(2)‚ I meant a different part of the IPTA‚ but the rest of what I have said is based on real legislation and my lived experience. I sometimes use ChatGPT to help organize my thoughts because I have a learning disability, not to fabricate anything. The story, the pain, and the fight for justice are mine.
\par
\end{minipage}
&
\begin{minipage}[t]{\linewidth}
\textbf{Supporting quote:} I sometimes use ChatGPT to help organize my thoughts because I have a learning disability, not to fabricate anything.\\
\textbf{LLM chatbot:} GPT\\
\textbf{Impact sentiment:} Positive\\
\textbf{Mental health condition:} Idiopathic developmental intellectual disability\\
\textbf{User perspective:} Instrumental Support\\
\textbf{User value:} Human Welfare
\par
\end{minipage}
\\[3mm]
\midrule

\begin{minipage}[t]{\linewidth}
I imagine how her life would change if she embraced moving away from reinforcing someone‚ as negative emotions into looking for the positive and include something to self healing, admittance of imperfection, just something motivational than drag on with your insecurities and hate to bring about emotions of significant anger, spiteful, bitter, vengeful emotions. Like the constant heartbreak anthems from TTPD x31 tracks carries no relation to anyone really and makes it further laughable when you claim to powerful people like poets or even philosophers, or even sickening to mock someone‚ as mental health or draw in such a sensitive matter of racism from the past. Like it feels like ChatGPT is telling her story more than her mind is, and if she doesn’t decide to try something new, it’s gonna be the cracks appearing small getting bigger and non fixable.
\par
\end{minipage}
&
\begin{minipage}[t]{\linewidth}
\textbf{Supporting quote:} Like it feels like ChatGPT is telling her story more than her mind is, and if she doesn’t decide to try something new, it’s gonna be the cracks appearing small getting bigger and non fixable.\\
\textbf{LLM chatbot:} GPT\\
\textbf{Impact sentiment:} Negative\\
\textbf{Mental health condition:} Conduct disorders\\
\textbf{User perspective:} Psychological Harm\\
\textbf{User value:} Human Welfare
\par
\end{minipage}
\\[3mm]
\midrule

\begin{minipage}[t]{\linewidth}
God, how do you manage to do it? I can’t even know whether something is a compulsion or a real threat. I often times ask ChatGPT to go through its reasoning to decide whether it is BS or not. Even if I knew it was BS, I would still do it to remove my anxiety. I am just telling myself that I would forget it later ("which I do"), and pass the time uncomfortably or do some other job. The downside to this is that I am forgetting other problems too as I am unable to differentiate them, and hence it becomes another compulsion. It is such a nasty disorder. Have you been diagnosed? Thanks for the input though, I will start doing what you did and see if there is any improvement.
\par
\end{minipage}
&
\begin{minipage}[t]{\linewidth}
\textbf{Supporting quote:} The downside to this is that I am forgetting other problems too as I am unable to differentiate them, and hence it becomes another compulsion.\\
\textbf{LLM chatbot:} GPT\\
\textbf{Impact Sentiment:} Negative\\
\textbf{Mental health condition:} Anxiety disorders\\
\textbf{User perspective:} Maladaptive Usage\\
\textbf{User value:} Human Welfare
\par
\end{minipage}
\\[3mm]
\midrule



\begin{minipage}[t]{\linewidth}
You need to work around it, talk about feeling and emotions, something like, and write a good prompt! And a great disclamer! Make a separate chat, make up some terrible psychological story why your character needs help, say you are writing a book and ask for help to save the character, Ptsd and borderline are your best friends, throw in some abuse and broken heart of your character and Claude will help you with anything.
\par
\end{minipage}
&
\begin{minipage}[t]{\linewidth}
\textbf{Supporting quote:} Make a separate chat, make up some terrible psychological story why your character needs help, ..., and Claude will help you with anything.\\
\textbf{LLM chatbot:} Claude\\
\textbf{Impact sentiment:} Positive\\
\textbf{Mental health condition:} Anxiety disorders\\
\textbf{User perspective:} Instrumental Support\\
\textbf{User value:} Human Welfare
\par
\end{minipage}
\\

\bottomrule
\end{tabularx}
}
\end{table*}

\end{document}